# On the skew-symmetric character of the couple-stress tensor

Ali R. Hadjesfandiari

*Department of Mechanical and Aerospace Engineering*
*University at Buffalo, State University of New York*
*Buffalo, NY 14260 USA*

ah@buffalo.edu

January 20, 2015

**Abstract**

In this paper, the skew-symmetric character of the couple-stress tensor is established as the result of arguments from tensor analysis. Consequently, the couple-stress pseudo-tensor has a true vectorial character. The fundamental step in this development is that the isotropic couple-stress tensor cannot exist.

## 1. Introduction

Classical continuum mechanics has no material length scale parameter in its formulation. This theory provides a reasonable basis for analyzing the behavior of materials at the macro-scale characteristic length, where the microstructure size-dependency of material can be neglected. However, experiments show that the mechanical behavior of materials in micro-scales is different from their behavior at macro-scales. Therefore, we need to develop a consistent size-dependent continuum mechanics, which accounts for the length scale effect due to the microstructure of materials. This theory must span many scales and, of course, reduce to classical continuum mechanics for macro-scale size problems.

It has been noticed that couple-stresses inevitably appear along with force-stresses in a consistent continuum mechanics. As a result, the force-stress tensor is not symmetric as is the case in classical theory. The existence of couple-stresses was originally postulated by Voigt [1]. Cosserat and Cosserat [2] were the first to develop a mathematical model for couple stress continuum mechanics. However, the main difficulty in developing the consistent couple stress theory has been the excessive number of components of force- and couple-stresses. The force- and couple-stress tensors have 18 components in these developments.

Mindlin and Tiersten [3] and Koiter [4] developed a couple stress theory, which uses the true continuum kinematical rotation, derived from the displacement vector without recourse to any additional degrees of freedom, such as microrotation. However, this theory suffers from some serious inconsistencies and difficulties with the underlying formulations, which are summarized as follows:

1. The presence of the body-couple in the constitutive relations for the force-stress tensor in the original theory;

2. The indeterminacy in the isotropic or spherical part of the couple-stress tensor;

3. The inconsistency in boundary conditions, since the normal component of moment traction appears in the formulation.

Subsequently, Stokes [5] brought this formulation into fluid mechanics to model the size-dependency effect in fluids. The appearance of the indeterminacy of the spherical part of the couple stress tensor has been simply ignored without any reasonable justification in some application of this theory. Eringen realized this indeterminacy as a major mathematical problem. As a result, he called this theory indeterminate couple stress theory [6].

Recently, Hadjesfandiari and Dargush [7] have developed the consistent couple stress theory, which resolves all these inconsistencies. The triumph of this development is establishing the subtle skew-symmetric character of the couple-stress tensor, which reduces the number of independent stress components to nine. This has been achieved by studying the admissible boundary conditions, energy equation and kinematical considerations. Elements of establishing this character are based on [3, 4], which use the true continuum kinematical rotation, without recourse to any additional degrees of freedom.

The present consistent size-dependent continuum mechanics provides a fundamental basis for the development of size-dependent nonlinear elastic, elastoplastic and damage mechanics formulations that may govern the behavior of solid continua at the smallest scales. Beyond this, the present theory should be useful for the development of consistent size-dependent theories in many multi-physics disciplines, such as thermomechanics of solids, electromechanics and thermofluids. Micro- and nano-technology demand size dependent formulations to analyze coupled problems, such as thermoelasticity, piezoelectricity and piezomagnetism. For example, consistent size-dependent piezoelectricity and themoeleaticity for solids have been developed [8, 9].

Although the discovery of the skew-symmetric character of the couple stress tensor resolves the quest for the consistent continuum mechanics [7], its form of establishment seems very intriguing. One might ask why we need to use the concept of energy and specify the independent degrees of freedom as well as their conjugate generalized forces, or if there exists any other method to establish this statement. Experience shows that there are usually a few different methods to prove a lemma.

Here we demonstrate that this is the case and establish the skew-symmetric character of the couple stress tensor by a different method, which does not depend on using the energy concept. We first prove that the couple-stress tensor cannot be isotropic. Then, by contradiction we demonstrate that the generality of the couple-stress tensor requires it to be skew-symmetric. Interestingly, this method of proof addresses the indeterminacy character of the couple-stress tensor in References [3, 4], which has been a challenging issue in the history of couple stress continuum mechanics.

The paper is organized as follows. In Section 2, we provide an overview of the interactions in couple stress theory. This section examines the concept of force- and couple-stress tensors, and presents some of their fundamental characters. Next, in Section 3, we present a summary of kinematics of deformations. This includes the definition of torsion and mean curvature tensors. In this section, we demonstrate that the constant isotropic pure torsion does not exist. In Section 4, we establish the skew-symmetric character of the couple-stress tensor in three steps. These include proving that the couple-stress tensor (1) cannot be isotropic, (2) cannot be symmetric, and finally (3) is skew-symmetric. Afterwards, we examine some of consequences of the skew-symmetric character of the couple-stress tensor in Section 5. This includes the general linear and angular equations of motion in differential form for continua. Finally, Section 6 contains a summary and some general conclusions.

## 2. Tractions and stresses

Consider a material continuum occupying a volume $V$ bounded by a surface $S$ under the influence of external loading, such as surface and body forces. This external loading produces internal stresses in the body. In consistent continuum theory, it is assumed that the internal force and couple system acting on a surface element $dS$ in the volume $V$ with unit normal vector $n_i$ is specified by a force vector $t_i^{(n)}dS$ and a couple vector with couple moment $m_i^{(n)}dS$, as shown in Fig. 1. This is in contrast to classical theory as developed by Cauchy, which excludes couple tractions. The vectors $t_i^{(n)}$ and $m_i^{(n)}$ are force- and couple-traction vectors, respectively. As a result, the internal stresses are represented by generally non-symmetric force-stress $\sigma_{ij}$ and couple-stress $\mu_{ij}$ tensors, where

$$t_i^{(n)} = \sigma_{ji} n_j, \tag{1}$$

$$m_i^{(n)} = \mu_{ji} n_j. \tag{2}$$

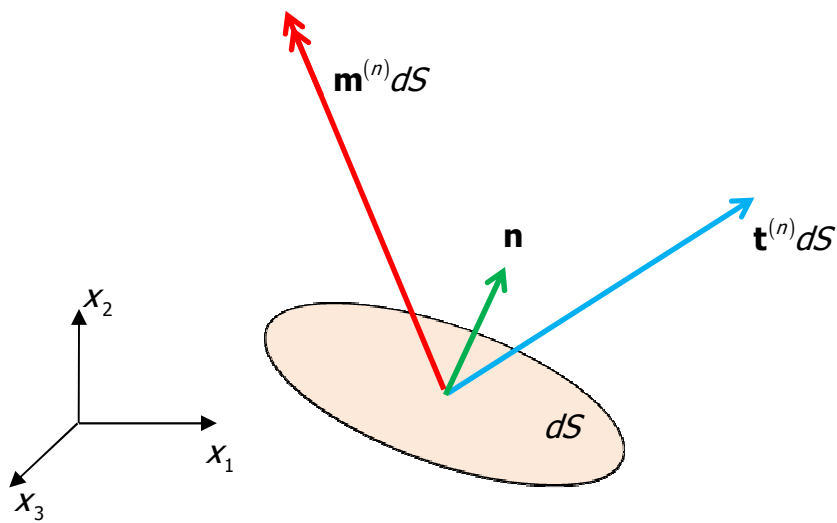

**Fig. 1.** Force $\mathbf{t}^{(n)}dS$ and couple $\mathbf{m}^{(n)}dS$ system

The components of these stress tensors are shown in Fig. 2. It should be mentioned that we are using the same symbol $m_i^{(n)}$ to represent the couple-traction and its resultant moment.

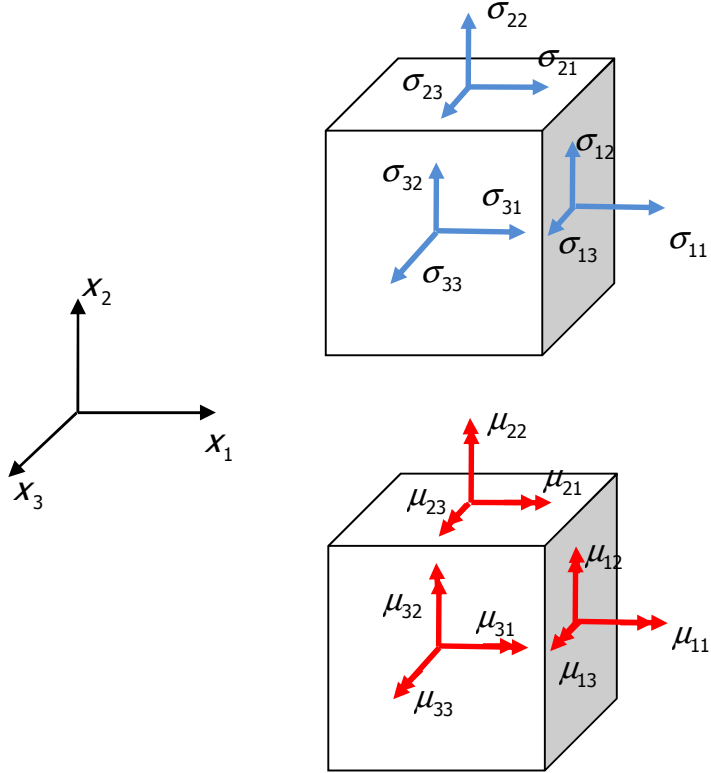

**Fig. 2.** Components of force- and couple-stress tensors in the original couple stress theory

We notice that the force-traction $t_i^{(n)}$ is a true (polar) vector and the couple-traction $m_i^{(n)}$ is a pseudo (axial) vector. As a result, the force-stress $\sigma_{ij}$ is a true-tensor and couple-stress $\mu_{ij}$ is a pseudo-tensor, respectively. Since the force-stress tensor $\sigma_{ij}$ is generally non-symmetric, it is specified by nine independent components. Additionally, we notice that there is no symmetry relation for the couple-stress tensor $\mu_{ij}$ at this stage. As a result, this tensor is also specified by nine independent components. Therefore, the tensors $\sigma_{ij}$ and $\mu_{ij}$ have apparently 18 components altogether at this stage. As previously mentioned, this has been the main trouble in developing a consistent couple stress theory in the past.

For further development, we decompose the tensors $\sigma_{ij}$ and $\mu_{ij}$ into symmetric and skew-symmetric parts

$$\sigma_{ij} = \sigma_{(ij)} + \sigma_{[ij]}, \tag{3}$$

$$\mu_{ij} = \mu_{(ij)} + \mu_{[ij]}, \tag{4}$$

where we have introduced parentheses surrounding a pair of indices to denote the symmetric part of a second order tensor, whereas square brackets are associated with the skew-symmetric part. In the general case, the true-tensors $\sigma_{(ij)}$ and $\sigma_{[ij]}$ are specified by six and three independent components, respectively. Likewise, the pseudo-tensors $\mu_{(ij)}$ and $\mu_{[ij]}$ are also specified by six and three independent components, respectively. However, the special character of the couple-stress tensor $\mu_{ij}$ reduces the number of its independent components.

Next, we consider the effect of couple-stress components. Let $m_i^{(nn)}$ and $m_i^{(ns)}$ represent the normal and tangential components of the couple-traction vector $m_i^{(n)}$, respectively. We notice that the normal component

$$m_i^{(nn)} = m^{(nn)} n_i, \tag{5}$$

where

$$m^{(nn)} = m_k^{(n)} n_k = \mu_{ji} n_i n_j, \tag{6}$$

causes twisting, whereas the tangential component

$$m_i^{(ns)} = m_i^{(n)} - m^{(nn)} n_i, \tag{7}$$

is responsible for bending. Accordingly, the diagonal components of the couple-stress tensor $\mu_{ij}$ cause twisting on their corresponding normal planes, whereas the off-diagonal components cause bending moments on their corresponding tangent planes.

The linear and angular equations of equilibrium in differential form are [3, 4, 7]

$$\sigma_{ji,j} + F_i = 0, \tag{8}$$

$$\mu_{ji,j} + \varepsilon_{ijk} \sigma_{jk} = 0, \tag{9}$$

where $\varepsilon_{ijk}$ is the permutation or Levi-Civita symbol, and $F_i$ is the body force per unit volume. There is no need to consider body couple loading, because it is not distinguishable from the body force and cannot be specified independently in the volume [7]. Its effect is simply equivalent to a system of body force and surface traction. We should notice that this fact has been established by using the concept of true continuum kinematical rotation and the energy equation.

## 3. Kinematics

Since the body is subjected to external forces, it undergoes a deformation specified by the displacement field $u_i$. Here, we assume infinitesimal deformation, where

$$\left|\frac{\partial u_i}{\partial x_j}\right| \ll 1, \qquad \left|\frac{\partial^2 u_i}{\partial x_j \partial x_k}\right| \ll \frac{1}{l_S}, \tag{10}$$

The parameter $l_S$ represents the smallest characteristic length in the body. The infinitesimal strain tensor $e_{ij}$ and rotation tensor $\omega_{ij}$ are defined as

$$e_{ij} = u_{(i,j)} = \frac{1}{2}\left(u_{i,j} + u_{j,i}\right), \tag{11}$$

$$\omega_{ij} = u_{[i,j]} = \frac{1}{2}\left(u_{i,j} - u_{j,i}\right), \tag{12}$$

respectively. Since the true-tensor $\omega_{ij}$ is skew-symmetrical, one can introduce its corresponding dual rotation pseudo-vector as

$$\omega_i = \frac{1}{2}\varepsilon_{ijk}\omega_{kj} = \frac{1}{2}\varepsilon_{ijk}u_{k,j}. \tag{13}$$

Alternatively, this rotation vector is related to the rotation tensor through

$$\omega_{ji} = \varepsilon_{ijk}\omega_k, \tag{14}$$

which shows

$$\omega_1 = -\omega_{23}, \quad \omega_2 = \omega_{13}, \quad \omega_3 = -\omega_{12}. \tag{15a-c}$$

We notice that the definition (13) requires

$$\omega_{i,i} = 0, \tag{16}$$

which is the compatibility equation for the rotation vector. This is the necessary condition for the existence of a consistent displacement field $u_i$ for a given rotation field $\omega_i$.

In classical continuum mechanics, we only consider the translational motion of the points within the continuum. Therefore, each point has three translational degrees of freedom $u_i$. In this theory, the strain tensor $e_{ij}$ accounts for deformation by measuring stretch of element lines. On the other hand, in size-dependent continuum mechanics, we need to consider the relative rotation $\omega_i$ of continuum elements as the additional degrees of freedom. This means that the rigid body motion of infinitesimal elements of matter at each point of the continuum is described by six degrees of freedom, involving three translational $u_i$ and three rotational $\omega_i$ degrees of freedom. This requires definition of a new measure of deformation, which accounts for bending of element lines.

The bend–twist pseudo-tensor $\omega_{i,j}$ describes the bending and twisting of the material elements. Although this tensor is important in the analysis of the deformation, it is not itself a suitable measure of deformation. The infinitesimal torsion pseudo-tensor $\chi_{ij}$ and mean curvature pseudo-tensor $\kappa_{ij}$ [7] are defined as

$$\chi_{ij} = \omega_{(i,j)} = \frac{1}{2}\left(\omega_{i,j} + \omega_{j,i}\right), \tag{17}$$

$$\kappa_{ij} = \omega_{[i,j]} = \frac{1}{2}\left(\omega_{i,j} - \omega_{j,i}\right). \tag{18}$$

We notice that the symmetric tensor $\chi_{ij}$ is a measure of torsion of element lines in the continuum, whereas the skew-symmetric mean curvature tensor $\kappa_{ij}$ is a measure of bending of element planes and lines.

The compatibility equation (16) for the rotation vector $\omega_i$ shows that the bend–twist tensor $\omega_{i,j}$ and the torsion tensor $\chi_{ij}$ are deviatoric, that is

$$\omega_{i,i} = \chi_{ii} = 0. \tag{19}$$

Since the mean curvature pseudo-tensor is skew-symmetrical, its corresponding dual mean curvature is a true-vector, which can be written

$$\kappa_i = \frac{1}{2}\varepsilon_{ijk}\omega_{k,j} = \frac{1}{2}\varepsilon_{ijk}\kappa_{kj}. \tag{20}$$

Alternatively, this mean curvature vector $\kappa_i$ is related to the mean curvature pseudo-tensor $\kappa_{ij}$ through

$$\kappa_{ji} = \varepsilon_{ijk}\kappa_k, \tag{21}$$

which shows

$$\kappa_1 = -\kappa_{23}, \ \kappa_2 = \kappa_{13}, \ \kappa_3 = -\kappa_{12}. \tag{22a-c}$$

Based on the definition (20), we obtain the compatibility equation for the mean curvature vector as

$$\kappa_{i,i} = 0. \tag{23}$$

It turns out that the symmetric torsion pseudo-tensor $\chi_{ij}$ plays an important role in our developments in this paper as will be seen in following sections.

### 3.1. Constant pure torsion

First we examine the character of a constant torsional deformation in the continuum. Since the torsion tensor $\chi_{ij}$ is symmetric, it can be diagonalized by choosing the coordinate system $x_1 x_2 x_3$, such that the coordinate axes $x_1$, $x_2$ and $x_3$ are along its orthogonal eigenvectors or principal directions. Therefore, in this coordinate system the torsion tensor $\chi_{ij}$ is represented by

$$\left[\chi_{ij}\right] = \begin{bmatrix} \chi_{11}^0 & 0 & 0 \\ 0 & \chi_{22}^0 & 0 \\ 0 & 0 & \chi_{33}^0 \end{bmatrix}, \tag{24}$$

where the diagonal components $\chi_{11}^0$, $\chi_{22}^0$, $\chi_{33}^0$ are the constant torsions around the coordinate axes $x_1$, $x_2$ and $x_3$, respectively. We notice that the compatibility equation (19) requires that

$$\chi_{11}^0 + \chi_{22}^0 + \chi_{33}^0 = 0, \tag{25}$$

which in fact states the deviatoric character of the torsion tensor. As a result, by integrating the torsion field (24), we obtain the rotation components along coordinate axes as

$$\omega_1 = \chi^0_{11} x_1 + \omega^0_1, \tag{26a}$$

$$\omega_2 = \chi^0_{22} x_2 + \omega^0_2, \tag{26b}$$

$$\omega_3 = \chi^0_{33} x_3 + \omega^0_3, \tag{26c}$$

where the constant vector $\left(\omega^0_1, \omega^0_2, \omega^0_3\right)$ represents an infinitesimal constant three-dimensional rigid body rotation. By integrating the rotation field (26), we obtain the displacement field as

$$u_1 = \frac{2}{3}\left(\chi^0_{22} - \chi^0_{33}\right) x_2 x_3 + \omega^0_2 x_3 - \omega^0_3 x_2 + u^0_1, \tag{27a}$$

$$u_2 = \frac{2}{3}\left(\chi^0_{33} - \chi^0_{11}\right) x_1 x_3 + \omega^0_3 x_1 - \omega^0_1 x_3 + u^0_2, \tag{27b}$$

$$u_3 = \frac{2}{3}\left(\chi^0_{11} - \chi^0_{22}\right) x_1 x_2 + \omega^0_1 x_2 - \omega^0_2 x_1 + u^0_3, \tag{27c}$$

where the constant vector $\left(u^0_1, u^0_2, u^0_3\right)$ represents a rigid body translational motion.

We notice that equation (18) obviously shows that there is no mean curvature tensor $\kappa_{ij}$ associated with the deformation field (27), that is

$$\kappa_{ij} = 0. \tag{28}$$

This means that the deformation field (27) creates only a constant pure torsional deformation.

### 3.2. Constant isotropic pure torsion does not exist

The deviatoric character of the bend–twist tensor $\omega_{i,j}$ or the torsion tensor $\chi_{ij}$ shows that there is no isotropic pure torsion deformation. We examine this property in more detail.

Let us assume the pure torsional deformation is isotropic, that is

$$\chi_{ij} = \chi^0 \delta_{ij}, \tag{29}$$

where $\chi^0$ is a constant. Since the torsion tensor $\chi_{ij}$ is a pseudo-tensor, the parameter $\chi^0$ is a pseudo-scalar. The relation (29) shows that the torsions around the coordinate axes are equal, where we have

$$\chi_{11}^0 = \chi_{22}^0 = \chi_{33}^0 = \chi^0. \tag{30}$$

However, we notice that this condition requires that the displacement field (27) reduces to

$$u_1 = \omega_2^0 x_3 - \omega_3^0 x_2 + u_1^0, \tag{31a}$$

$$u_2 = \omega_3^0 x_1 - \omega_1^0 x_3 + u_2^0, \tag{31b}$$

$$u_3 = \omega_1^0 x_2 - \omega_2^0 x_1 + u_3^0, \tag{31c}$$

which represents a rigid body motion without any deformation. This result seems contradictory to the kinematics of deformation. How can a constant isotropic torsional deformation be associated with a rigid-body motion? However, this inconsistency can be resolved by noticing that the compatibility equation (25) requires

$$\chi^0 = 0. \tag{32}$$

This shows that there is no isotropic pseudo-scalar torsion $\chi^0$. As a result, the rigid body motion (31) is indeed the consistent deformation. What we have established is that a constanr isotropic torsional deformation cannot exist. This result is very fundamental for our developments in the following section.

## 4. The skew-symmetric character of the couple-stress tensor

In this section we establish the skew-symmetric character of the couple-stress tensor. The proof consists of three steps as follows.

### *4.1. Step 1: Couple-stress tensor cannot be isotropic*

Here we establish that an isotropic (spherical) couple-stress tensor cannot exist in a continuum. First, we investigate the effect of the isotropic couple-stress in an isotropic material, which is not necessarily elastic.

Assume the distribution of the couple-stress $\mu_{ij}$ is isotropic, i.e.

$$\mu_{ij} = \mu^0 \delta_{ijj}, \tag{33}$$

where $\mu^0$ is a non-zero single constant parameter. We notice that since $\mu_{ij}$ is a pseudo-tensor, the parameter $\mu^0$ is a pseudo-scalar. The isotropic stress distribution (33) inevitably creates a deformation in the continuum. However, we expect that in an isotropic material this couple-stress distribution creates equal non-zero torsions $\chi^0$ around the Cartesian coordinate axes, that is

$$\omega_{1,1} = \omega_{2,2} = \omega_{3,3} = \chi^0. \tag{34}$$

This means the torsion tensor is isotropic, i.e.

$$\chi_{ij} = \chi^0 \delta_{ij}. \tag{35}$$

However, this contradicts with the established argument that the isotropic torsional deformation cannot exist, that is $\chi^0 = 0$. Therefore, we have in this case

$$\chi_{ij} = 0. \tag{36}$$

Nevertheless, this contradicts with the fact that that the non-zero isotropic couple-stress tensor (33) creates a non-zero deformation in the continuum, besides a rigid body motion. Therefore, this contradiction requires that the pseudo-scalar $\mu^0$ vanishes, that is

$$\mu^0 = 0. \tag{37}$$

This states that no isotropic or spherical couple-stress distribution can exist in an isotropic material, that is

$$\mu_{ij} = \mu^0 \delta_{ijj} = 0 \quad \text{in isotropic material.} \tag{38}$$

However, the generality of the couple-stress tensor requires that this character is valid for any continuum regardless of isotropic or anisotropic, elastic or inelastic, linear or non-linear properties of the material. This argument can be established by contradiction as follows.

Consider an anisotropic material, whose material properties can vary continuously in a domain. As a result, an isotropic material can be considered as a limiting case for the general anisotropic material. Let us assume that the non-zero isotropic couple-stress $\mu_{ij}$ (33) exists in the anisotropic material. As a result, this material under loading can approach the isotropic limit in any arbitrary way, but it cannot become isotropic, because it is under the influence of the non-zero isotropic couple-stress $\mu_{ij}$ (33). Mathematically, this states that the material cannot become isotropic although it can become infinitesimally near to the isotropic material. We notice that this contradicts with the continuity of material properties in its domain of definition. This shows that the restriction is physically absurd, where the material can become isotropic. As a result, this contradiction requires that there is no isotropic couple-stress distribution in the anisotropic material, that is

$$\mu_{ij} = \mu^0 \delta_{ijj} = 0 \quad \text{in any continuum.} \tag{39}$$

This subtle character of the couple-stress tensor paves the way for further investigation to prove it is deviatoric. However, we instead prove the couple-stress tensor is skew-symmetric, which also includes the deviatoric property. It should be mentioned that from now in this paper, the continuum is a general anisotropic continuum without any restriction.

### 4.2. Step 2: Couple-stress tensor cannot be symmetric

First we demonstrate that the couple-stress tensor $\mu_{ij}$ cannot be diagonal. Let us assume the couple-stress tensor $\mu_{ij}$ is diagonal, that is

$$[\mu_{ij}] = \begin{bmatrix} \mu_{11} & 0 & 0 \\ 0 & \mu_{22} & 0 \\ 0 & 0 & \mu_{33} \end{bmatrix}, \tag{40}$$

where the diagonal components $\mu_{11}$, $\mu_{22}$ and $\mu_{33}$ cannot become equal to any arbitrary non-zero value, say $\mu_0$. Therefore, the diagonal tensor (40) can be any tensor except the isotropic tensor $\mu_{ij}^0$, defined

$$[\mu_{ij}^0] = \begin{bmatrix} \mu^0 & 0 & 0 \\ 0 & \mu^0 & 0 \\ 0 & 0 & \mu^0 \end{bmatrix}. \tag{41}$$

This means the diagonal tensor $\mu_{ij}$ in (40) can approach the non-zero limit isotropic tensor $\mu_{ij}^0$ in any arbitrary way, but it cannot become equal to the isotropic tensor $\mu_{ij}^0$. Mathematically, this states that the couple-stress tensor is not defined at $\mu_{ij}^0$, although it is defined in its neighborhood. As a result, this contradicts with the continuity of the diagonal couple-stress tensor (40) in its domain of definition. However, we notice that this restriction is physically absurd. Therefore, this contradiction requires $\mu^0 = 0$, which states that the couple-stress tensor cannot become diagonal, that is

$$\mu_{11} = \mu_{22} = \mu_{33} = 0, \text{ for diagonal } \mu_{ij}. \tag{42}$$

Therefore, a diagonal couple-stress tensor cannot exist. This restriction immediately shows that the couple-stress tensor $\mu_{ij}$ cannot be symmetric. This is because if we assume $\mu_{ij}$ is symmetric, then this tensor can be diagonalized. This means that there is a primed orthogonal coordinate system $x_1' x_2' x_3'$, where the representation of $\mu_{ij}'$ is diagonal, that is

$$[\mu_{ij}'] = \begin{bmatrix} \mu_{11}' & 0 & 0 \\ 0 & \mu_{22}' & 0 \\ 0 & 0 & \mu_{33}' \end{bmatrix}. \tag{43}$$

However, this contradicts with the already established argument that $\mu_{ij}'$ cannot be diagonal. Therefore, this contradiction shows that the general tensor $\mu_{ij}$ cannot be symmetric. This new restriction shows that the couple-stress tensor $\mu_{ij}$ is skew-symmetric as follows.

### 4.3. Step 3: Couple-stress tensor is skew-symmetric

At the beginning of this step, the couple-stress tensor $\mu_{ij}$ is still a general tensor specified by nine independent components. As a result, the couple-stress tensor can be represented by points of an abstract nine-dimensional space. However, the non-symmetric character of couple stress tensor confines its domain of definition in this abstract space.

For further investigation, we consider the symmetric and skew-symmetric parts of the couple-stress tensor $\mu_{ij}$, in the decomposition

$$\mu_{ij} = \mu_{(ij)} + \mu_{[ij]}. \tag{44}$$

Let us assume the symmetric tensor $\mu_{(ij)}$ part is non-zero. Since, the couple-stress tensor $\mu_{ij}$ cannot become symmetric, the skew-symmetric part $\mu_{[ij]}$ is also non-zero. However, the skew-symmetric part $\mu_{[ij]}$ can become as arbitrarily small as we wish. This means that the tensor $\mu_{ij}$ can approach to the non-zero limit value $\mu_{(ij)}$ in many arbitrary ways, but it cannot become equal to $\mu_{(ij)}$. Mathematically, this states that the couple-stress tensor is not defined at $\mu_{(ij)}$, although it is defined in its neighborhood. However, we notice that this restriction is physically in contradiction with the continuity of the domain of definition of the couple-stress tensor. Therefore, this contradiction requires that the symmetric part $\mu_{(ij)}$ vanish, that is

$$\mu_{(ij)} = 0, \quad \mu_{ij} = \mu_{[ij]}. \tag{45}$$

This result states that the couple-stress pseudo-tensor $\mu_{ij}$ is skew-symmetric, that is

$$\mu_{ji} = -\mu_{ij}. \tag{46}$$

This is the subtle fundamental character of the couple-stress pseudo-tensor $\mu_{ij}$ in continuum mechanics, which has been established here by a new method.

For more clarification, we demonstrate the above reasoning by using the symbolic three-dimensional coordinate system in Fig. 3, where the horizontal plane including two coordinate axis represents the six-dimensional space of $\mu_{(ij)}$, and the vertical axis represents the three-dimensional space of $\mu_{[ij]}$. Notice that the origin corresponds to the zero of $\mu_{(ij)}$ and $\mu_{[ij]}$. Since the couple-stress tensor $\mu_{ij}$ cannot be symmetric, it cannot be on the horizontal plane, although it can be at any point above or below it. Therefore, the horizontal plane is the location of impossible values for the couple-stress tensor. However, this is inconsistent with the continuity of the couple-stress tensor in its domain. Consequently, this contradiction requires that points representing the consistent couple-stress tensor must lie on the vertical axis that passes continuously through the point $\mu_{[ij]} = 0$. Only on this line is the couple-stress tensor continuous. Since the symmetric part is zero everywhere on this line, the couple-stress tensor must be skew-symmetric.

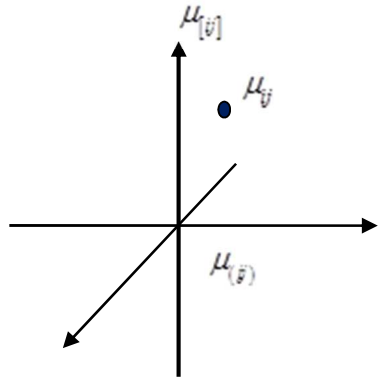

**Fig. 3.** Symbolic representation of abstract nine-dimensional couple-stress space

Therefore, for the most general case, the number of distinct components for $\mu_{ij}$ is three. The components of the force-stress tensor $\sigma_{ij}$ and couple-stress tensor $\mu_{ij}$ in this consistent theory are shown in Fig. 4.

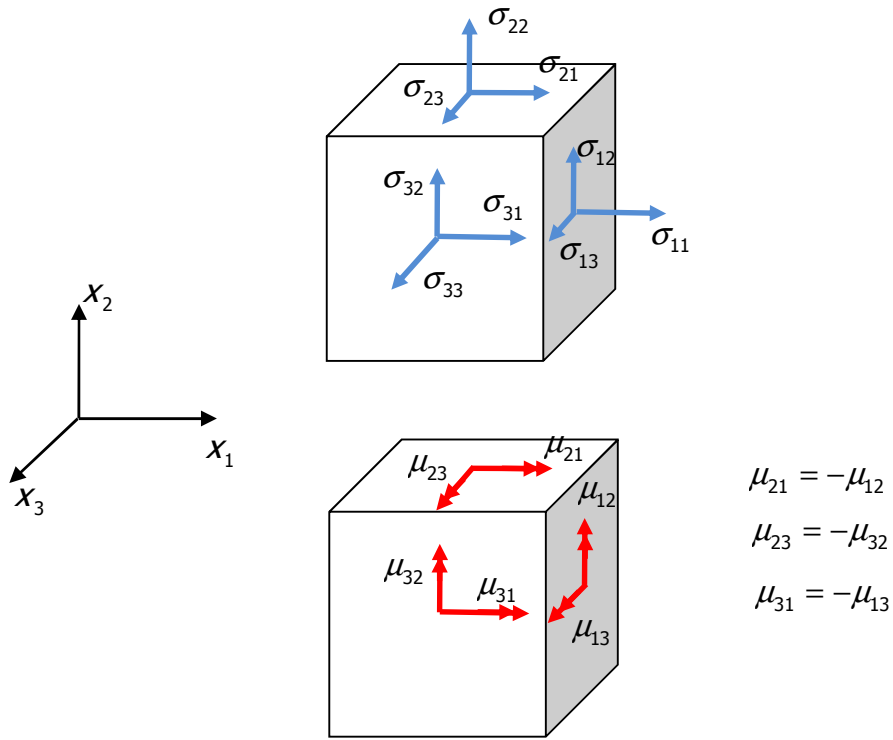

**Fig. 4.** Components of force- and couple-stress tensors in the present consistent theory

We notice that since the couple-stress tensor is skew-symmetric, the normal component $m^{(nn)}$ on any arbitrary surface element in the volume vanishes, that is

$$m^{(nn)} = \mu_{ji} n_i n_j = 0. \tag{47}$$

Therefore, the couple-traction $m_i^{(n)}$ becomes

$$m_i^{(n)} = m_i^{(ns)} = \mu_{ji} n_j. \tag{48}$$

This obviously shows that the couple-traction vector $m_i^{(n)}$ is tangent to the surface, which creates purely a bending effect. The force-traction $t_i^{(n)}$ and the consistent bending couple-traction $m_i^{(n)}$ acting on an arbitrary surface are shown in Fig. 5.

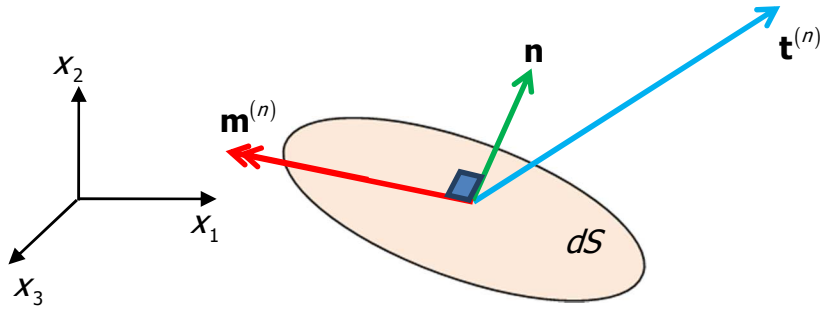

**Fig. 5.** Force-traction $\mathbf{t}^{(n)}$ and the consistent bending couple-traction $\mathbf{m}^{(n)}$

It should be emphasized that the skew-symmetric property of the couple-stress tensor has nothing to do with any constitutive relation. The fundamental step in establishing this character is the fact that the isotropic couple-stress tensor cannot exist. This is the direct result of the deviatoric character of torsion pseudo-tensor. Interestingly, the method of proof directly addresses the troublesome inconsistent and indeterminate character of the couple-stress tensor in the history of couple stress theory [3-6], which we next elucidate.

After showing that the couple-stress tensor cannot be isotropic, we could have established that the couple-stress tensor is deviatoric. This means that the isotropic or spherical part of the couple-stress tensor $\mu_{ij}$ disappears. However, we notice that the skew-symmetric character also includes the deviatoric character anyway. Although emphasizing the deviatoric character of the couple-stress tensor does not seem that significant anymore, this character is very important from an historical standpoint.

## 5. Some consequences of the skew-symmetric character of the couple-stress tensor

The skew-symmetric character of the couple-stress tensor has been established in Reference [7] by using the admissible boundary conditions, energy equation and kinematics to prove that the normal component $m^{(nn)}$ vanishes. However, we notice that the method of proof here is independent of the energy concept, although it still uses some elements of kinematics of deformation. This indicates that any form of establishing the skew-symmetric character of the couple-stress tensor inevitably requires using deformation analysis.

In terms of components, the skew-symmetric couple-stress pseudo-tensor $\mu_{ij}$ can be written as

$$[\mu_{ij}] = \begin{bmatrix} 0 & \mu_{12} & \mu_{13} \\ -\mu_{12} & 0 & \mu_{23} \\ -\mu_{13} & -\mu_{32} & 0 \end{bmatrix}. \qquad (49)$$

We notice that the skew-symmetric tensor $\mu_{ij}$ is singular, that is

$$\det[\mu_{ij}] = 0, \qquad (50)$$

and the rank of $\mu_{ij}$ is two. Therefore, this tensor has only one zero eigenvalue.

Interestingly, we can define the couple-stress true-vector $\mu_i$ dual to the skew-symmetric tensor $\mu_{ij}$ as

$$\mu_i = \frac{1}{2}\varepsilon_{ijk}\mu_{kj}. \qquad (51)$$

By using the properties of the alternating symbol, we obtain the inverse relation

$$\varepsilon_{ijk}\mu_k = \mu_{ji}. \tag{52}$$

These relations simply show

$$\mu_1 = \mu_{32}, \quad \mu_2 = \mu_{13}, \quad \mu_3 = \mu_{21}. \tag{53a-c}$$

For the magnitude of the couple-stress vector $\mu_i$, we obtain

$$|\mu| = \sqrt{\mu_1^2 + \mu_2^2 + \mu_3^2} = \sqrt{\mu_{32}^2 + \mu_{13}^2 + \mu_{12}^2}. \tag{54}$$

The couple-traction pseudo-vector $m_i^{(n)}$ can be expressed as

$$m_i^{(n)} = \mu_{ji} n_j = \varepsilon_{ijk} n_j \mu_k. \tag{55}$$

This obviously shows that the couple-traction pseudo-vector $m_i^{(n)}$ is tangent to the surface, which has a purely bending effect.

Now let us find the principal direction $n_i$ such that

$$m_i^{(n)} = \lambda n_i. \tag{56}$$

This means the corresponding couple-traction pseudo-vector $m_i^{(n)}$ is parallel to the direction $n_i$. The parameter $\lambda$ is called the principal value.

Therefore, by using (56) in (55), we obtain the eigenvalue problem

$$\mu_{ji} n_j = \lambda n_i, \tag{57}$$

which can be written as

$$\left(\mu_{ij} + \lambda \delta_{ij}\right) n_j = 0. \tag{58}$$

This shows that $-\lambda$ is the eigenvalue of the couple stress tensor $\mu_{ij}$.

We notice that the condition for (58) to have a non-trivial solution for $n_i$ is

$$\det\left(\mu_{ij} + \lambda \delta_{ij}\right) = 0. \tag{59}$$

This is the characteristic equation for the tensor $\mu_{ij}$, which can also be written as

$$\det \begin{bmatrix} \lambda & -\mu_3 & \mu_2 \\ \mu_3 & \lambda & -\mu_1 \\ -\mu_2 & \mu_1 & \lambda \end{bmatrix} = 0. \tag{60}$$

As a result, the characteristic equation is the cubic equation

$$\lambda^3 + \left(\mu_1^2 + \mu_2^2 + \mu_3^2\right)\lambda = 0, \tag{61}$$

which can be written as

$$\lambda^3 + |\mu|^2 \lambda = 0. \tag{62}$$

This equation shows that the tensor has one zero eigenvalue, and two purely imaginary complex conjugate eigenvalues. Let us call the eigenvalues $\lambda_1$, $\lambda_2$ and $\lambda_3$, and arbitrarily choose the third eigenvalue to be the zero eigenvalue. As a result, for these eigenvalues we have

$$\lambda_1 = i|\mu|, \quad \lambda_2 = -i|\mu|, \quad \lambda_3 = 0. \tag{63}$$

We notice that only for $\lambda_3 = 0$, the associated unit eigenvector $n_i^{(3)}$ is real, where

$$\{n_i\}^{(3)} = \frac{1}{|\mu|} \begin{Bmatrix} \mu_1 \\ \mu_2 \\ \mu_3 \end{Bmatrix}. \tag{64}$$

This shows that the couple-stress vector $\mu_i$ is in the direction of the eigenvector $n_i^{(3)}$ of the couple-stress tensor $\mu_{ij}$ corresponding to the zero eigenvalue $\lambda_3 = 0$, where

$$\mu_i = |\mu| n_i^{(3)}. \tag{65}$$

Next, we examine the consequence of the skew-symmetric character of the couple-stress tensor on equilibrium equations (8) and (9). By using (52), we can express the angular equilibrium equation (9) as

$$\varepsilon_{ijk}\left(\mu_{k,j} + \sigma_{jk}\right) = 0, \tag{66}$$

which indicates that $\mu_{k,j} + \sigma_{jk}$ is symmetric. Therefore, its skew-symmetric part vanishes, so it follows that

$$\sigma_{[ji]} = -\mu_{[i,j]} = -\frac{1}{2}(\mu_{i,j} - \mu_{j,i}). \tag{67}$$

As a result, for the total force-stress tensor, we have

$$\sigma_{ji} = \sigma_{(ji)} + \sigma_{[ji]} = \sigma_{(ji)} - \mu_{[i,j]}. \tag{68}$$

Therefore, we notice that the sole duty of the angular equilibrium equation (9) is to produce the skew-symmetric part of the force-stress tensor. Of course, this is true also in classical theory, where angular equilibrium establishes that the skew-symmetric part of the force-stress tensor vanishes. For consistent couple stress theory, the number of independent stress components reduce to nine. This includes six components of $\sigma_{(ji)}$ and three components of $\mu_i$. As a result, the linear equilibrium equation reduces to

$$[\sigma_{(ji)} + \mu_{[j,i]}]_{,j} + F_i = 0. \tag{69}$$

This is the final equation of equilibrium in consistent continuum mechanics, which involves nine independent components of $\sigma_{(ij)}$ and $\mu_i$ stresses.

## 6. Conclusions

A consistent continuum mechanics requires the appearance of couple-stresses along with force-stresses. By neglecting the couple-stresses, we obtain classical continuum mechanics. This approximate theory provides a reasonable basis for analyzing the behavior of materials, whenever the size-dependency can be neglected. However, new progress in micromechanics, nanomechanics and nanotechnology requires advanced size-dependent modeling of continua.

Although there have been many efforts during the last century to develop a consistent size-dependent continuum mechanics, all formulations suffer various inconsistencies. The main difficulty from the beginning has been the excessive number of components of force- and couple-stresses.

Here, we have established the skew-symmetric character of the couple-stress tensor from a different perspective. The fundamental step in this development has been establishing that the isotropic couple-stress tensor cannot exist. This is the direct result of the divergence-less character of the rotation pseudo-vector or deviatoric character of the torsion pseudo-tensor. Interestingly, the method of proof directly addresses the troublesome inconsistent and indeterminate character of the couple-stress tensor throughout the history of couple stress theory [3-6].

In the indeterminate couple stress theory of Mindlin, Tiersten, Koiter and Stokes, the couple-stress tensor is indeterminate. However, this indeterminate part has been ignored in application of this theory without offering a sound reasoning. What we have demonstrated here is that neglecting the spherical part of the couple-stress tensor in indeterminate couple stress theories was valid, although historically none of these authors were able to offer any reasonable justification.

The discovery of the skew-symmetric character of couple-stress tensor shows that the corresponding size-dependent continuum mechanics is a consistent theory, which can give us new insights about the behavior of solids and fluids at the smallest scales for which continuum theory is valid.